\documentclass[10pt, conference, compsocconf]{IEEEtran}
\usepackage{amsmath}
\usepackage{dsfont}
\usepackage{algorithm}
\usepackage{algorithmic}
\usepackage{array}
\usepackage{cite}
\ifCLASSINFOpdf
   \usepackage[pdftex]{graphicx}
\else
\fi

\newcommand{\remove}[1]{}
\renewcommand{\baselinestretch}{0.99}
\newcommand{\cent}{\mathrm{cr}}
\newcommand{\ncent}{\mathrm{ncr}}

\newcommand{\pr}{\mathrm{pr}}
\newcommand{\out}{\mathrm{out}}
\newtheorem{theorem}{Theorem}
\newtheorem{lemma}[theorem]{Lemma}
\newcommand{\C}{\mathrm{\mathcal{X}_c}} 
\newcommand{\N}{\mathrm{\mathcal{X}_n}} 
\newcommand{\WC}{\mathrm{\mathcal{W}_c}}
\newcommand{\WN}{\mathrm{\mathcal{W}_n}}
\newcommand{\F}{\mathrm{\mathcal{C}}}
\newcommand{\FN}{\mathrm{\mathcal{N}}}
\begin{document}

\title{Non-Conservative Diffusion and its Application to Social Network Analysis}
\author{
\IEEEauthorblockN{Rumi Ghosh, Kristina Lerman, Tawan Surachawala}\\
     \IEEEauthorblockA{USC Information Sciences Institute\\
    Marina del Rey, CA 90292, USA\\
       \{rumig,lerman,tawans\}@isi.edu}
       \and
\IEEEauthorblockN{Konstantin Voevodski}\\
      \IEEEauthorblockA{Department of Computer Science\\
       Boston University\\
       kvodski@gmail.com}
       \and
\IEEEauthorblockN{ Shanghua Teng}\\
       \IEEEauthorblockA{University of Southern California\\
Los Angeles, CA 90007, USA\\
   shanghua@usc.edu}}

\date{}

\maketitle
\begin{abstract}
Is the random walk appropriate for modeling and analyzing social processes? We argue that many interesting social phenomena, including epidemics and information diffusion, cannot be modeled as a random walk, but instead must be modeled as broadcast-based or non-conservative diffusion. To produce meaningful results, social network analysis algorithms have to take into account differences between these diffusion processes.
We formulate conservative (random walk-based) and non-conservative (broadcast-based) diffusion mathematically and show how these are related to well-known metrics: PageRank and Alpha-Centrality respectively.  This formulation allows us to unify two distinct areas of network analysis --- centrality and epidemic models --- and leads to insights into the relationship between diffusion and network structure, specifically, the existence of an epidemic threshold in non-conservative diffusion. We demonstrate, by ranking nodes in an online social network used for broadcasting news, that non-conservative Alpha-Centrality leads to a better agreement with empirical ranking schemes than conservative PageRank.
In addition, we give a scalable approximate algorithm for computing the Alpha-Centrality in a massive graph. 
We hope that our investigation will inspire further exploration of the applications of non-conservative diffusion in social network analysis.
\end{abstract}


\IEEEpeerreviewmaketitle

\section{Introduction}
Social network analysis algorithms examine the topology of the network to identify central nodes within it or groups of tightly connected nodes. In many cases, these algorithms make implicit assumptions about the underlying diffusion process taking place on the network~\cite{Borgatti05}. Some of the best-known algorithms used for graph partitioning~\cite{approximatePR} and ranking, including PageRank and its variants~\cite{PageRank,Jeh03}, are based on the random walk~\cite{Tong06,Fortunato07}. A random walk on a graph is a stochastic process which starts at some node, and at each time step randomly selects one of the neighbors of the current node.
The random walk is used to model chemical diffusion and other physical processes in which the total amount of the diffusing substance remains constant. 
However, the random walk may not be appropriate for modeling phenomena of greatest interest to social scientists, including adoption of innovation~\cite{Rogers03,Bettencourt05}, the spread of epidemics~\cite{Anderson91,Hethcote00} and word-of-mouth recommendations~\cite{Goldenberg01}, viral marketing campaigns~\cite{Kempe03,Iribarren09}, and information diffusion~\cite{Lerman10icwsm}. These examples are modeled as contact processes, where an activated or ``infected'' node activates its neighbors with some probability. Rather than picking one of the neighbors, in these stochastic processes each node \emph{broadcasts} to all its neighbors. Therefore, unlike the random walk, which conserves the amount of substance diffusing on the network, contact processes are fundamentally non-conservative. When an idea, information, or disease spreads from one individual to her neighbors, the amount of information or disease changes (Chapter 5, \cite{NewmanBarabasiWattsBook}). 
If the random walk cannot model these social processes, can we trust results of social network analysis algorithms that are based on the random walk? If not, what are the appropriate metrics and methods to use for network analysis? And how can we empirically evaluate their performance?

In this paper we present a mathematical formulation of conservative and non-conservative diffusion and demonstrate how these are related to two well-known centrality metrics used to rank nodes in a network: PageRank~\cite{PageRank} and Alpha-Centrality~\cite{Bonacich87}. While PageRank is known to be equivalent to conservative diffusion~\cite{Tong06,Fortunato07}, we show that Alpha-Centrality is related to non-conservative diffusion, of which epidemic models are the best known example. Our formulation unifies two distinct research areas within network analysis --- centrality measures and epidemic models --- and leads to insights into relationship between dynamic processes and network structure. One consequence of the analysis is the existence of a threshold, called epidemic threshold~\cite{Wang03}, below which non-conservative diffusion dies out, but above which it reaches significant fraction of nodes within the network. We elucidate connection between the properties of Alpha-Centrality and the location of the epidemic threshold.

We demonstrate empirically that the choice of the centrality metric impacts our ability to identify central or influential nodes within a network. Specifically,
we study online social network of Digg involved in spreading news stories. The spread of news on Digg can be modeled as an epidemic process~\cite{Versteeg11}, and hence represents non-conservative diffusion. One benefit of using social media data sets is that user activity on these sites provides an independent measure of influence. We define two empirical measures of influence that serve as the ground truth for ranking users within this social network. We compare the rankings produced by different centrality metrics to the ground truth and show that non-conservative Alpha-Centrality leads to a better agreement with the ground truth than conservative PageRank.
Finally, we present an approximate algorithm that can efficiently compute Alpha-Centrality for massive graphs and give a proof of its performance guarantees.

Specifically, the paper makes the following contributions:
\begin{itemize}
\item Define and classify diffusion processes occurring on networks (Section \ref{sec:classes}).
\item Establish a connection between  diffusion and network structure  (Section \ref{sec:structure}). 
We also show how centrality metrics are related to diffusion processes occurring on the network. 
\item Empirically validate the hypothesis that non-conservative metric better predicts central people in an online social network used for (non-conservative) information diffusion than a conservative metric (Section~\ref{sec:OSN}).
\item Provide a fast approximate algorithm to compute Alpha-Centrality (Section~\ref{sec:fast}).
\end{itemize}
\remove
{
\begin{itemize}
\item Mathematically formulate conservative and non-conservative diffusion and relate these to PageRank and Alpha-Centrality (Section~\ref{sec:diffusion}) and elucidate the relationship between epidemics and network structure (Sec.~\ref{sec:threshold})
\item Elucidate the link between diffusion models and centrality and empirically validate centrality metrics on online social networks data (Sec.~\ref{sec:centrality})
\item Provide a fast approximate algorithm to compute Alpha-Centrality (Sec.~\ref{sec:fast})
\end{itemize}
}

\section{Classes of Diffusion Processes}
\label{sec:classes}
We represent a network by a directed, weighted graph $G = (V,E)$ with $V$ nodes and $E$ edges.
We use $w[u,v]$ to specify the weight of the edge from $u$ to $v$.  The adjacency matrix of the graph is defined as: $A[u,v]= w[u,v]$ if $(u,v) \in E$; otherwise, $A[u,v]= 0$. $N(u)$ is the set of  out-neighbors of $u$: $N(u) = \lbrace v \in V \vert (u,v) \in E \rbrace$,  $d_{\out}(u)$ is the out-degree of $u$: $d_{\out}(u) = \sum_{v \in N(u)} w[u,v]$, and $d_{\max}$ is the maximum out-degree of any node in the graph. Note that the $L_1$-norm of any argument is given by $||.||_1$ . 

{\em Network diffusion} is a dynamic stochastic process that distributes some quantity, which we generically refer to as weight, on a network or a graph.  Diffusion process is described mathematically by a function $F: (R^+ \cup \{0\})^{|V|} \rightarrow (R^+ \cup \{0\})^{|V|}$, i.e., a map from a $|V|$-dimensional non-negative vector to a $|V|$-dimensional non-negative vector (here $V$ is the number of nodes).   The vector $x \in (R^+ \cup \{0\})^{|V|}$  represents the weight each node  has at time $t$. The function $F(x)$ maps the weight vector at time $t$ to the weight vector at time $t+1$.

\subsection{Conservative Diffusion}
We call a stochastic process $\F_t: (R^+ \cup \{0\})^{|V|} \rightarrow (R^+ \cup \{0\})^{|V|}$ that simply redistributes the weights among the nodes of the graph, with the total weight remaining constant, \emph{conservative diffusion}. In other words, in conservative diffusion 
for all $x \in (R^+ \cup \{0\})^{|V|}$, $||x||_1 = ||\F_t(x)||_1$. 

To motivate our mathematical formulation of conservative diffusion, we imagine a hypothetical society where each member has some amount of money to redistribute. If money cannot be created or destroyed, money redistribution represents a conservative diffusion process. Let $\C(t)$ be the vector representing the amount of money each member has at time $t$, and $\Delta(t)$ represent the amount they {receive} at time $t$. We consider a distribution process where the amount \emph{redistributed} at each step, depends on the money each member \emph{received} in the previous step.  We focus on this redistribution process, because, as we show later, this is the process underlying  popular network models. A different conservative process could be one in which the amount redistributed in each step depends on the amount each member \emph{had} in the previous time step. This would lead to a different mathematical formulation of the diffusion process.

 At time $t+1$, each member retains a fraction $(1-\alpha)$, with $0 \le \alpha \le1$, of this amount  and distributes the rest among its neighbors.  Let $\WC$ be the \emph{transfer matrix}, with $\WC[p,q]$ representing the fraction of the amount to be redistributed by node $p$ transferred to $q$. Therefore, the amount of money nodes receive  at time $t+1$ via redistribution can be written as:
$$\Delta(t+1)=\alpha \Delta(t)\WC.$$
Thus the transfer matrix encodes the rules of diffusion. If each member  \emph{divides} $\alpha \Delta(t)$ equally amongst her out-neighbors, then $\WC=D^{-1}A$, where the degree matrix $D$ is a diagonal matrix of out-degrees, and $A$ is the adjacency matrix. 

Step by step, conservative diffusion looks as follows. Initially, at time $t=0$, let the weight each node receives be $\Delta(0)=\C(0)$.   Let the process begin at  time $t=1$, when each node keeps  $(1-\alpha)$ of that amount and divides the rest ($\alpha\Delta(0)$) evenly between its out-neighbors.  The amount that out-neighbors receive from redistribution at time $t=1$ is  $\Delta(1) =\alpha \Delta(0)\WC=\alpha \C(0)\WC$. 

At time $t=2$, each node retains $(1-\alpha)$ of the amount $\Delta(1)$ it received at time $t=1$, and divides the rest among its out-neighbors. Therefore, the amount  received by the out-neighbors is
$\Delta(2)= \alpha \Delta(1)\WC=\alpha^2 \C(0)\WC^2$. 

Continuing with this process further,  at any time $t>0$, each nodes  retains $(1-\alpha)$ of the amount of it received at time $t-1$,
\begin{eqnarray}
(1-\alpha)\Delta(t-1) &= &(1-\alpha)\alpha\Delta(t-2)\WC  \nonumber\\
&=& (1-\alpha)\alpha^{t-1} \C(0) \WC^{t-1},
\label{eq:Cons7} 
\end{eqnarray}
\noindent and divides the rest among her out-neighbors. 
Hence,  the amount  received by the out-neighbors is 
\begin{equation}
\Delta(t) = \alpha\Delta(t-1)\WC= \alpha^t \C(0)\WC^t.
\label{eq:Cons100} 
\end{equation}

The total weight (or amount of money in our example) the nodes have  at time $t$, $\C(t)$, is the amount they retained from all previous time steps and the amount they receive from in-neighbors at time $t$:
\begin{eqnarray}
\C(t) &= & (1-\alpha)\sum_{k=0}^{t-1}\Delta(k) + \Delta(t)  \nonumber \\
                  &= &\sum_{k=0}^{t-1}(1-\alpha) \alpha^{k} \C(0) \WC^{k}+\alpha^{t} \C(0) \WC^{t} \nonumber \\
 &= &(1-\alpha)\C(0)+\alpha \C(t-1)\WC.
                  \label{eq:Cons8}
\end{eqnarray}

\noindent As $t\to\infty$, this equation reduces to
\begin{eqnarray}
\C(t\to\infty) &=&(1-\alpha)\C(0)+\alpha \C(t\to\infty)\WC \nonumber\\
&=&(1-\alpha)\C(0){(I-\alpha \WC)}^{-1}
  \label{eq:Cons10}
\end{eqnarray}

The transfer matrix $\WC$ is a stochastic matrix, since its rows sum up to 1. If, as described above, the weight to be redistributed at each step is divided equally between the out-neighbors, then $\WC=D^{-1}A$. 
However, if instead each node decides to keep a portion $\delta$ of this amount, this leads to a more general form of the transfer matrix: 
\begin{equation}
\WC= \delta I+(1-\delta) D^{-1} A.
\label{eq:Wcons}
\end{equation}


Note that in our hypothetical society, the total amount of money remains constant: if $\F_t: \C(0) \rightarrow \C(t)$ defines a diffusion process, then $||\C(0)||_1 = || \F_t(\C(0))||_1$. Hence this is a \emph {conservative diffusion process}.
In the above scenario, $\F_t$ is a linear mapping; therefore, we call the diffusion processes given by Eqs.~\ref{eq:Cons8} and \ref{eq:Cons10}  \emph{linear conservative diffusion}.  In a more general representation, $\F_t$ can even be a non-linear mapping, describing non-linear conservative diffusion.

\subsubsection*{Random Walk as Conservative Diffusion}
Like money transfer, a {random walk} on a graph can be modeled as a conservative diffusion process, since the probability to find a random walker on any node of the graph is always one. A random walk with random jumps or restarts can be described mathematically as follows. Let the initial probability to find  the random walker on any node be uniform, i.e., $\C(0)[i]=\frac{1}{|V|}$. At  any time $t$, with probability $\alpha$ the random walker at node $i$ chooses one of the neighbors of $i$ uniformly at random and jumps to it. With probability $(1-\alpha)$, it chooses any node on the graph uniformly at random and jumps to it. Let matrix $X$ encode the probability of jumping to any node, $X[i,j]=\frac{1}{|V|}$, and $\WC=D^{-1}A$.
Then the probability of finding the random walker at node $j$ at time $t$ is given by 
\begin{eqnarray*}
\C(t)& =& (1-\alpha)\C(t-1)X+\alpha \C(t-1)\WC \\
&=& (1-\alpha)\C(0)+\alpha \C(t-1)\WC.
\end{eqnarray*}
This is exactly the same as Eq.~\ref{eq:Cons8}. Therefore,  a random walk with a uniform starting vector is mathematically equivalent to a linear conservative diffusion process.

\subsection{Non-Conservative Diffusion}
A diffusion process where the total weight  can change in time is a \emph{non-conservative diffusion process}. Formally, a function $\FN_t: (R^+ \cup
\{0\})^{|V|} \rightarrow (R^+ \cup \{0\})^{|V|}$ defines a \emph{ non-conservative diffusion} process if for some $x \in (R^+ \cup \{0\})^{|V|}$, $||x||_1 \neq ||\FN_t(x)||_1$.

To illustrate the difference between conservative and non-conservative processes, we return to our hypothetical society. Again, imagine that each member  has some amount of money, however, unlike the previous example, each member also has a money printing machine, so that instead of dividing the money she receives equally between her out-neighbors, she can give each neighbor the same amount by printing extra money as needed. 


Let $\Delta(t)$ be the vector representing the amount of money each member receives at time $t$. At the next time step, each member prints a fraction $\alpha$ of this amount to give to each of her out-neighbors. The  additional amount that she produces for her out-neighbors  can be expressed using the \emph{replication matrix} ${\WN}=A$. Therefore, $\Delta(t+1)=\alpha \Delta(t)\WN$.

Initially, let $\Delta(0)=\N(0)$. At time $t=1$, each member prints $\alpha  \Delta(0)$ for each of her out-neighbors:
$$\Delta(1)=\alpha  \Delta(0) \WN  = \alpha \N(0) \WN.$$
Similarly, at time $t=2$, 
$$\Delta(2)= \alpha  \Delta(1) \WN = \alpha^2  \N(0) \WN^2.$$
Continuing this process,  additional amount of money each member  produces or receives at time $t$ is:
\begin{equation}
\Delta(t) =  \alpha \Delta(t-1) \WN= \alpha^t \N(0) \WN^t
\label{eq:NonCons1}
\end{equation}
\noindent 
Therefore, the total amount that each member has at time $t$ is obtained by summing up  the additional amount she accrues
or receives from her in-neighbors at each time step:
\begin{eqnarray}
\N(t)&=& \sum^{t}_{k=0} \Delta(k)=\sum_{k=0}^{t} \N(0) (\alpha\WN)^k \nonumber \\
 &=&\N(0)+\alpha\N(t-1)  \WN
\label{eq:NonCons2}
\end{eqnarray}
\noindent 
 At time $t\to\infty$, Eq.~\ref{eq:NonCons2} reduces to
\begin{equation}
\N(t\to\infty) = \N(0){\sum_{k=0}^{t\to\infty}  (\alpha \WN)^k}
 \label{eq:NonCons4}
\end{equation}
\noindent
which can be solved to yield
\begin{eqnarray}
\N(t\to\infty) &=& \N(0)+ \N(t\to\infty)(\alpha \WN) \nonumber \\
&=&{\N(0)}{(I-\alpha \WN)}^{-1}.
 \label{eq:NonCons5}
\end{eqnarray}
This expression is defined for  ${\alpha}<1/{\lambda_1}$, where $\lambda_1$ is the largest eigenvalue, or spectral radius, of $\WN$.

More generally, if along with producing $\alpha$ of what it receives from each of its in-neighbors, a node also produces a portion $\delta$ of this amount for itself, this results in a more general form of the replication  matrix: 
\begin{equation}
\WN= \frac{\delta}{\alpha} I+A.
\label{eq:Wnoncons}
\end{equation}
\remove
{
Equation \ref{eq:NonCons5} reduces to 
\begin{equation}
\N(t\to\infty) ={\N(0)}{((1-\alpha \delta) I -\alpha A)}^{-1}.
 \label{eq:NonCons6}
\end{equation}
}
\remove{
Here, the replication matrix is \emph{static} for all $t$, $\WN=A$. If the replication matrix changes with time, $\WN(t)$, the amount of money each member has at time $t$ is: 
\begin{equation}
\N(t)= \N(0) +\N(0)\sum^{t}_{j=1} \prod_{k=1}^{j}(\delta I+\alpha \WN(k)) 
\label{eq:NonCons3}
\end{equation}
}

The diffusion process defined by Eqns.~\ref{eq:NonCons2}--\ref{eq:NonCons5} is non-conservative, since $||\N(0)||_1 \neq || \FN_t(\N(0))||_1$. Moreover, it is linear, although the function  $\FN_t$ may also be non-linear.

We can model non-conservative diffusion as a random walk with birth, where at each time step, the random walker gives birth to one or more new walkers.  The number of random walkers on the network, therefore, will change with time. Several social phenomena can be modeled using this framework. In rumor propagation, for example, some information spreads in a community as people pass it to their neighbors. This process is non-conservative, since the number of informed individuals grows in time. We can model rumor propagation as a random walk on the friendship graph, where the random walker (rumor) randomly selects one of the neighbors of the informed node to move to, while leaving a clone of itself at the node. Cloning is required for the node to remain informed. If the informed node immediately forgot the rumor (no cloning required), than rumor propagation could be modeled by a simple random walk and would be conservative in nature, since the number of informed individuals would always be one. 
\remove
{
As another example, consider the spread of a disease within a community. At each time step, an infected person can transmit the virus to one or more of her neighbors and the number of viruses in the system does not remain constant. We model this as non-conservative diffusion process where a random walker clones itself at each time step, with each clone initiating its own random walk. 
}


\subsubsection*{Epidemics as Non-Conservative Diffusion}
\label{sec:threshold}
Non-conservative diffusion provides a useful framework for thinking about epidemics and other spreading processes and leads to insights into the relation between network structure and dynamics of spreading processes.  In a spreading process, information or virus spreads from an informed or infected individual to her network neighbors. In order to model a spreading process accurately, the structure of the underlying network has to be taken into account. Wang \emph{et al.}~\cite{Wang03} modified existing SIS models~\cite{Bailey:1975} to take network structure into account in order to describe the spread of epidemics in real networks.  We demonstrate that this model  is equivalent to the  \emph{linear non-conservative diffusion process} (Equation \ref{eq:NonCons2}). 

Consider a virus spreading on a network, where at each time step, a node infected with the virus may infect its out-neighbors with probability $\mu$ (virus birth rate). At each time step, an infected node may also be cured with probability $\beta$ (virus curing rate). Wang et al.~\cite{Wang03} showed that the probability $p_{i,t}$ that  node $i$ is infected at time $t$  can be written in matrix notation as 
$$
P_t=P_{t-1}((1-\beta) I+\mu A) =P_0((1-\beta)I+\mu A)^t
$$
\noindent where $P_t$ is a vector $(p_{1,t},\ p_{2,t},\ \ldots)$, and $P_0$ is the initial probability of infection.\footnote{This model holds true only when $p_{i,t}$ is very small and there may be situations where $p_{i,t}>1$. Therefore a more accurate interpretation is that the probability of infection is proportional to $p_{i,t}$.} This formulation makes the probability of infection at time $t$, $P_t$, exactly equal to the additional weight, $\Delta(t)$, accrued at each step in non-conservative diffusion,  as shown in Eq.~\ref{eq:NonCons1} with the replication matrix $\WN= \frac{1-\beta}{\mu} I+A$ and $\alpha=\mu$.
In the model described above, there exists an epidemic threshold $\tau$ such that for  $\mu/\beta < \tau$ epidemic will die out, and $\mu/\beta >\tau$ it will spread to a significant fraction of nodes~\cite{Wang03}. For any graph, this threshold is given by   the inverse of the largest eigenvalue of the  graph's adjacency matrix $A$: $\tau=1/|\lambda_1|$.

\section{Diffusion and Network Structure}
\label{sec:structure}

The complex interplay between network structure and diffusion has broad implications for modeling and understanding networks. While it is known that the macroscopic properties of diffusion (e.g., epidemic threshold) are affected by network structure~\cite{Wang03,Pastor-Satorras2001}, the impact of diffusion on our understanding of network structure is less appreciated. In this paper we show that social network analysis, specifically, identifying central or influential nodes, is affected by the characteristics of the diffusion process occurring on the network. Centrality metrics used for this task examine the topology of the network only. However, these metrics usually make implicit assumptions about the nature of diffusion process taking place on the network~\cite{Borgatti05}, with each metric leading to a different, even conflicting notion, of who the central nodes are. We show that the characteristics of network diffusion should be one of the guiding principles in choosing an appropriate network analysis algorithm.

\subsection {Centrality and Diffusion}
\label{sec:centdif}
A node's centrality predicts its relative importance, influence, or prestige within the network. Over the years many different centrality metrics have been introduced for social network analysis, including degree centrality, betweenness centrality~\cite{Freeman79}, eigenvector centrality~\cite{Bonacich01}, PageRank~\cite{PageRank} and Alpha-Centrality~\cite{bonacich72factoring}.  
 
\subsubsection{Page Rank} 
 A PageRank vector ${\pr}_{\alpha}(s,t)$ is the steady state probability distribution of a random walk with damping factor $\alpha$ (restart probability= $1-\alpha$).  The starting vector $s$, gives the probability distribution for where the walk transitions after restarting.  The transfer matrix encodes the transition probabilities of a random walk on the network, $W = D^{-1}A$. PageRank is the unique steady state solution ${\pr}_{\alpha}(s,\infty)$ of:
\begin{equation} 
\label{eq:pr}
{\pr}_{\alpha}(s,t) = (1-\alpha) s +  \alpha {\pr}_{\alpha}(s,t-1)W
\end{equation} 
For ease of convention, we denote PageRank by ${\pr}_{\alpha}(s)$. Hence
\begin{equation} 
\label{eq:pr1}
{\pr}_{\alpha}(s) = (1-\alpha) s +  \alpha {\pr}_{\alpha}(s)W
\end{equation} 

Equation~\ref{eq:pr1} is identical to the steady state solution of the linear conservative diffusion process given by Eq.~\ref{eq:Cons10}  
where $W=\WC=D^{-1}A$ and $s=\C(0)$. 
Therefore, \emph{PageRank is the steady state solution of conservative diffusion}, and PageRank is a {conservative metric}. 
Most of the other metrics derived from the random walk make an implicit assumption of  conservative diffusion taking place on a network. 

\subsubsection{Alpha-Centrality}
Alpha-Centrality~\cite{bonacich72factoring} measures the total number of paths from a node, exponentially attenuated by their length.
For a starting vector $s$ and attenuation parameter $\alpha$, the Alpha-Centrality vector is the steady state solution to:
\begin{equation}
\cent_{\alpha}(s,t) = s + \alpha \cent_{\alpha}(s,t-1)A.
\label{a-cen}
\end{equation}
The starting vector $s$ is usually taken as in-degree centrality~\cite{bonacich72factoring}. For ease of convention, we shall denote $\cent_{\alpha}(s, t\to \infty)$  by $\cent_{\alpha}(s)$. As $t \to \infty$, the solution converges to
\begin{equation}
\cent_{\alpha}(s)=s+ \alpha \cent_{\alpha}(s)A,
\label{a-cen1}
\end{equation}
\noindent which holds while $|\alpha| < \frac{1}{|\lambda_{1}|}$. 

One difficulty in applying Alpha-Centrality in network analysis is that its key parameter $\alpha$ is bounded by $\lambda_1$, the spectral radius of the network. As a result, the metric diverges at this value of the parameter. To overcome this,  \emph{normalized Alpha-Centrality} \cite{Ghosh11Physrev} has been recently introduced,  which we denote by $\ncent_{\alpha}(s,t)$. It normalizes the score of each node by the sum of the Alpha-Centrality scores of all the nodes. The new metric avoids the problem of bounded parameters while retaining the desirable characteristics of Alpha-Centrality, namely its ability to differentiate between local and global structures.

Normalized Alpha-Centrality $\ncent_{\alpha}(s,t \to \infty)$ is defined using the system of equations shown below:
\begin{equation}
\ncent_{\alpha}(s,t)  = \frac{1}{||\cent_{\alpha}(s,t)||_1} \cent_{\alpha}(s,t)
\label{n-a-cen-summation}
\end{equation}
\noindent The new metric is well defined for $\alpha \ge 0\ (\alpha \neq \frac{1}{|\lambda_1|})$.

Equation \ref{a-cen} and Eq.~\ref{n-a-cen-summation} are  mathematically equivalent to Eq.~\ref{eq:NonCons4}, 
with starting vector $\N(0)=c\cdot s$, where $c=1$ for Alpha-Centrality  and 
$$c= \frac{1}{\sum_{i,j}\sum_{k=0}^t \alpha^k A^k[i,j]}$$ 
\noindent for normalized Alpha-Centrality.  
Therefore, \emph{Alpha-Centrality is a steady state solution of linear non-conservative diffusion} and is a {non-conservative metric}. Other non-conservative metrics include degree centrality, Katz score~\cite{Katz53}, SenderRank~\cite{Kiss08}, and eigenvector centrality~\cite{Bonacich87}.

\subsection{Length Scales and Epidemic Threshold} 
The link between Alpha-Centrality (and normalized Alpha-Centrality) and non-conservative diffusion leads to a fundamental insight into  the relationship between network structure and the size of epidemics. Let us look more carefully at Equation \ref{eq:NonCons4}.
The weight distribution given by this equation depends on the initial weight distribution ($ \N(0)$) and the power series of matrices $S(\alpha,t)={\sum_{k=0}^{t}  (\alpha \WN)^k}$.   For illustrative purposes, we can interpret $ \WN$ to be the adjacency matrix of some graph $G'$. Then each element in the power series $S(\alpha,t)[i,j]$ can be interpreted as the number of attenuated paths from node $i$ to node $j$ up to length $t$ in that graph $G'$.  In Alpha-Centrality or normalized Alpha-Centrality, these paths determine the centrality of the node along with the initial distribution of weights. The probability of non-conservative diffusion reaching node $j$ from $i$ through a path of length $k$ is $\alpha^k$. $S(\alpha,t)[i,j]$ then characterizes the expected number times a non-conservative diffusion process initiated at node $i$ reaches node $j$ up until time $t$.  For example, let node $i$ be infected by a virus and initiate a viral infection in the network.   If viral infection can be modeled as linear non-conservative diffusion (Section \ref{sec:threshold}), the probability that node $j$ will get infected by the viral infection from node $i$ through a path of length $k$ would be $\alpha^k$. Then $S(\alpha,t)[i,j]$ would quantify the expected number of viruses reaching node $j$ when the viral infection is initiated at node $i$.

As shown in the Appendix, the expected path length of diffusion as $t\to \infty$, is $\frac{1}{1-\alpha\lambda_1}$ if $\alpha<\frac{1}{|\lambda_1| }$ and $O(t)$ if $\alpha >\frac{1}{|\lambda_1|}$. Therefore, $\frac{1}{|\lambda_1|}$ is a \emph{threshold}: for $\alpha $ below threshold, the expected path length converges with time, while for $\alpha$  above the threshold, it diverges.  Note that this threshold is equivalent to the epidemic threshold (Section \ref{sec:threshold}). Thus from the diffusion point of view, given the network structure and nature of diffusion,  $\alpha$ (for $\alpha <1/{\lambda_1}$) determines how far, on average, a node's effect will be felt and sets the length scale of the interaction. When $\alpha$ is small, Alpha-Centrality or normalized Alpha-Centrality probes only the local structure of the network. As $\alpha$ grows, structurally longer paths become more important, (normalized) Alpha-Centrality  becomes a global measure and  the weight diffuses to a greater number of nodes. 

\subsection{Choosing the Centrality Metric}
\begin{figure}[tbh]
\center{
\includegraphics[width=0.75\linewidth]{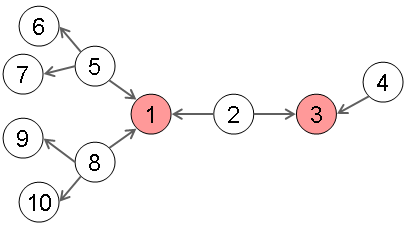}
}
\caption{ An example network, where node 1 has the highest Alpha-Centrality followed by node 3. In contrast node 3 has the highest PageRank followed by node 1. }
\label{fig:compare}
\end{figure}

When applied to the same network, different centrality metrics may lead to different, often incompatible, views of who the important nodes are. We illustrate these differences on a toy network shown in Fig.~\ref{fig:compare}, where a link from node $u$ to node $v$ indicates that  node $v$ is an out-neighbor of $u$, e.g., $u$ is a follower of $v$ in an online social network.

Even in this simple example, PageRank and Alpha-Centrality disagree about who the most important node is. PageRank without restarts ranks node 3 highest, followed by node 1. In contrast, Alpha-Centrality ranks node 1 above node 3.
The difference in rankings produced by the two centrality metrics is due to the difference in the underlying diffusion process that redistributes the weights of the nodes. Assume that all nodes start with equal weights, which then evolve according to the rules of diffusion.
In PageRank without restarts (damping factor $\alpha=1$), each follower divides its weight equally among its $d_{out}$ out-neighbors, and hence transfers a fraction $1/d_{out}$ to each. Thus, node 5 contributes 1/3 of its weight to node 1, and so will node 8. Node 3, on the other hand, will get the entire weight of node 4, giving it a higher weight than node 1 and therefore, a higher rank.

In contrast to PageRank, Alpha-Centrality has nodes update their weights by copying a portion of their followers' weights. For consistency with PageRank, we take $\alpha=1$. 
Thus, node 1 will receive the entire weights of nodes 2, 5 and 8, while node 3 will only receive the weights from nodes 2 and 4. Therefore, the weight of node 1 will be greater than node 3, and consequently, it will be ranked higher by Alpha-Centrality.

Which ranking is right? How do we choose the right centrality metric for our problem? We claim that the choice of the centrality metric has to be motivated by details of the diffusion process taking place on the network. To analyze networks on which processes such as random walk, web surfing, money and used goods exchange are taking place, conservative metrics, such as PageRank, are appropriate. On the other hand, to study social networks on which information or epidemics are spreading, non-conservative metrics, such as Alpha-Centrality, should be used. In other words, \emph{the centrality metric that best predicts important nodes in a network is one whose implicit dynamics most closely matches the diffusion process occurring on the network}.

\section{Predicting Influentials in Online Social Networks}
\label{sec:OSN}


Online social networks on sites such as Facebook, Twitter, and Digg have become important hubs of social activity and conduits of information. 
The ever-growing popularity of these networks and overwhelming amount of information contained in them, necessitates the need for a more principled approach to social network analysis and data mining. 
Correctly identifying influential nodes on these networks can have far-reaching consequences for identifying noteworthy content~\cite{Lerman10www}, targeted information dissemination~\cite{Kempe03}, and other applications. While a variety of methods~\cite{Cha10icwsm,Lee10www} have been used to identify influential users in online social networks, each metric leads to a different result, and no justification for these metrics have been proposed. 

Fortunately, by exposing activity of their users, online social networks provide a unique opportunity to study dynamic processes on networks. We analyze {information flow} on the social news aggregator Digg and use this data 
to empirically evaluate centrality metrics. By posting a story on Digg, submitter broadcasts it to her followers. When another user  votes for this story, she broadcasts it to her own followers. We claim that since broadcast-driven information diffusion on Digg is non-conservative in nature, a non-conservative metric will better identify influential users than a conservative metric.

The Digg dataset comprises around 300K users and over 1 million friendship links,  from which we can extract the directed follower network of active users. These users were active in spreading stories on Digg by either submitting them or voting for them, since both activities expose the story to the submitter or voter's followers. The data set contains more than 3 million votes on more than 3000 stories promoted to Digg's front page in June 2009.  Note that the underlying follower graph was extracted separately of user activity. In fact, user activity provides an independent measure of influence in online social networks that we use to evaluate the centrality metrics. 

\subsection{Empirical Estimates of Influence}
Katz and Lazarsfeld~\cite{Katz55} defined influentials as ``individuals who were likely to influence other persons in their immediate environment.'' In the years that followed, many attempts were made to identify people who influenced others to adopt a new practice or product by looking at how innovations or word-of-mouth recommendations spread~\cite{Brown87}.
The rise of online social networks has allowed researchers to trace the flow of information through social links on a massive scale. Using the new empirical foundation, some researchers proposed to measure a person's influence by the size of the cascade he or she triggers~\cite{Kempe03}. However, as Watts and Dodds~\cite{Watts07} note, ``the ability of any individual to trigger a cascade depends much more on the global structure of the influence network than on his or her personal degree of influence.''
Alternatively, Trusov \emph{et al.}~\cite{Trusov10} defined influential people in an online social network as those whose activity stimulates those connected to them to increase their activity, while Cha \emph{et al.}~\cite{Cha10icwsm} used the number of retweets and mentions to measure user influence on Twitter.

Motivated by these works, we measure influence by analyzing users' activity on an online social network. Suppose some user, the \emph{submitter}, posts a new story on Digg.   We measure the activity submitter's post generates by the number of times it is re-broadcast by followers. Whether or not a user will re-broadcast the story depends on  (\emph{i}) story quality and (\emph{ii}) influence of the submitter. We assume that story's quality is uncorrelated with the submitter.\footnote{This is a fairly strong assumption, but it appears to hold at least for Digg~\protect\cite{Lerman10www}.} Therefore, we can average out its contribution to the activity a submitter generates by aggregating over all stories submitted by the same user. We claim that the residual difference between submitters can be attributed to variations in influence. We propose two metrics to measure submitter's influence: (i) average number of follower votes her posts generate and (ii) average size of the cascades her posts trigger.

\subsection{Comparison of Centrality Metrics}
 \begin{figure}[tbh]
\begin{tabular}{c}
  \includegraphics[width=0.9\linewidth]{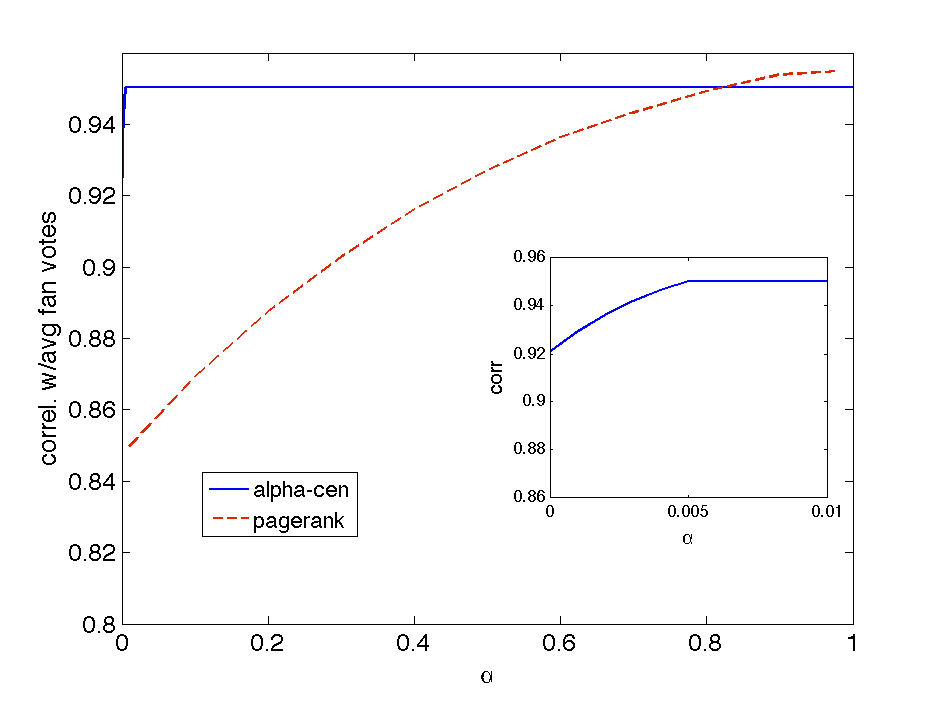} \\
  (a)\\
  \includegraphics[width=0.9\linewidth]{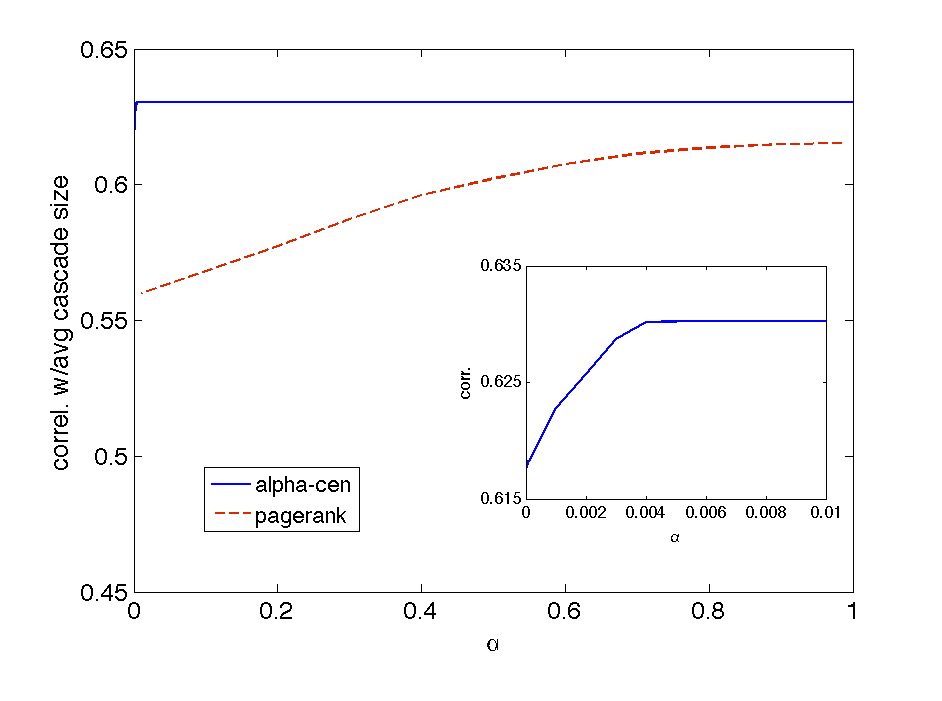} \\
  (b)
\end{tabular}
\caption{Correlation between the rankings produced by the empirical measures of influence and those predicted by normalized Alpha-Centrality and PageRank.  We use (a) the average number of follower votes and (b) average cascade size as the empirical measures of influence. The inset zooms into the variation in correlation for $0 \le \alpha \le 0.01$  }\label{fig:correl_digg}
\end{figure}

We use the empirical estimates of influence to rank a subset of users in our sample who submitted more than one story which received at least 100 votes. There were 289 Digg users in this sample. We used the rankings produced by either empirical estimate as the ground truth to evaluate the performance of different centrality metrics. We studied PageRank (with uniform starting vector) and normalized Alpha-Centrality, both of which were computed considering the entire Digg follower network as a graph, with  the thousands of users as nodes and the millions of friendship links as edges.  We use Pearson's correlation coefficient (since ties in rank may exist) to compare the rankings {predicted} by the different centrality metrics with the ground truth. Figure~\ref{fig:correl_digg} shows how the correlation in rankings changes with the parameter $0\le \alpha \le 1$. This parameter stands for the \emph{attenuation factor}  for normalized Alpha-Centrality (see Equation~\ref{a-cen}) and the \emph{damping factor} (\emph{restart probability}=$1-\alpha$) for PageRank  (see Equation~\ref{eq:pr}).  If we used Alpha-Centrality instead of normalized Alpha-Centrality, we would have been bounded by  its formalization, to compute  the rankings only for $\alpha <\frac{1}{|\lambda_1|}$. Note that the correlation of PageRank  at $\alpha=0$ (restart probability=1) with the empirical estimate cannot be computed because standard deviation of PageRank rankings would be zero in this case.
Various studies have tested different damping factors for Page Rank, but it is generally assumed that the damping factor should be set around $\alpha=0.85$ \cite{PageRank}. Boldi et al. \cite{Boldi:2005} claim that in case of PageRank, ``for real-world graphs values of $\alpha$ close to 1 do not give a more meaningful ranking.''  Except for values $\alpha$ close to 1, the influence rankings calculated from Alpha-Centrality correlated better with the empirical estimates of influence rankings than PageRank rankings.
Therefore, we conclude that Alpha-Centrality predicts central users in the Digg social network better than PageRank. 

\section{Approximation Algorithm for \\ Alpha-Centrality}
\label{sec:fast}
In order to compute the exact Alpha-Centrality vector we have to solve Equation \ref{a-cen}, which requires us to compute a matrix inverse.  Computing a matrix inverse in a naive implementation, takes $O(n^{3})$ time (where $n$ is the number of nodes in the network), so this is difficult to compute for large networks.  One way to compute an approximate solution is to use the alternate formulation given in Equation \ref{eq:NonCons2}, and compute $s(I + \alpha A + \alpha^{2} A^{2} + \alpha^{3} A ^{3} + \ldots)$, until the $\alpha^{i}$ coefficient grows sufficiently small.  While this technique is effective in practice, computing $A^{i}$ in each iteration, using a naive implementation would have must take at least $n^2$ time, and it is not clear how many iterations we need to get a good approximation.  In this section we present an algorithm for approximating Alpha-Centrality, which has a single parameter that controls both the runtime and the quality of the produced approximation.

\begin{algorithm}
\caption{Approximate-Centrality($V,E,s,\alpha,\delta$)}
\begin{algorithmic}[1]
\STATE $\epsilon = \delta || s || _{1} / n$;
\STATE $r = s$;
\STATE Queue q = new Queue();
\FOR{each $u \in V$}
\STATE $\tilde{cr}(u) = 0$;
\IF{$r(u) > \epsilon$}
\STATE q.add($u$);
\ENDIF
\ENDFOR
\WHILE{q.size() $>$ 0}
\STATE $u$ = q.dequeue();
\STATE $\tilde{cr}(u) = \tilde{cr}(u) + r(u)$;
\STATE $T = \alpha \cdot r(u)$;
\STATE $r(u) = 0$;
\FOR{each $v \in N(u)$}
\STATE $r(v) = r(v) + T \cdot w(u,v)$;
\IF{!q.contains($v$) and $r(v) > \epsilon$}
\STATE q.add($v$);
\ENDIF
\ENDFOR
\ENDWHILE
\RETURN $\tilde{cr}$;
\end{algorithmic}
\label{alg-main}
\end{algorithm}

A description of our algorithm is given in Algorithm~\ref{alg-main}.  Our procedure is similar to the algorithm for approximating PageRank that is given in \cite{approximatePR}.  Our algorithm takes the network, the starting vector $s$, $\alpha$, and an approximation parameter $\delta$ ($0 < \delta \le 1$) as input, and computes an approximate Alpha Centrality vector where each entry has error of at most $\delta$ (see Theorem~\ref{theorem:Correctness}).  In order to approximate a centrality vector with starting vector $s$, we maintain an \emph{approximate centrality} vector $\tilde{cr}$ and a \emph{residual} vector $r$.  Initially $r$ is equivalent to the starting vector $s$; the algorithm iteratively moves content from $r$ to $\tilde{cr}$ until each entry in $r$ is small.  

When the $\alpha$ parameter is fixed, we use $\cent(s)$ to denote $\cent_{\alpha}(s)$.  We will also use $\lbrack \cent(s) \rbrack (u)$ to refer to how much content vertex $u$ has in $\cent(s)$.  We  give our formal performance guarantee for Algorithm~\ref{alg-main} in Theorem \ref{theorem:Correctness}. This performance guarantee is based on Lemma \ref{lemma:LoopInvariant}, which shows that in any step of the algorithm, the approximate centrality computed for Alpha-Centrality with $s$ as starting vector, is always exactly equivalent to Alpha Centrality with $s-r$ as starting vector, where $r$ is the residual vector in that step i.e.  throughout the execution of the algorithm, the error in the approximate centrality vector is dependent on the amount of content remaining in the residual vector.

Our arguments depend on the linearity of the centrality computation with respect to the starting vector, which is easy to verify.  We can show that $\cent_{\alpha}(s_{1}) + \cent_{\alpha}(s_{2}) = \cent_{\alpha}(s_{1} + s_{2})$, and $c \cdot \cent_{\alpha}(s) = \cent_{\alpha}(c \cdot s)$.

\begin{lemma}\label{lemma:LoopInvariant}
The invariant $\tilde{cr} = \cent(s - r)$ is maintained throughout the execution of the while-loop.
\end{lemma}

\begin{proof}
Before the loop starts, we have $r = s$ and $\tilde{cr} = \vec{0}$, so $\cent(s-r) = \cent(\vec{0}) = \vec{0} = \tilde{cr}$.  We can also show that if $\tilde{cr} = \cent(s - r)$ holds prior to an iteration of the loop, then $\tilde{cr}' = \cent(s - r')$ is still true after the iteration, where $\tilde{cr}'$ and $r'$ are the updated approximate centrality and residual vectors.

We first observe that $\cent(s)A = \cent(sA)$.  To see this, consider that by definition $\cent(s) = s + \alpha \cdot \cent(s)A$.  Multiplying this equation by $A$ we get $\cent(s)A = sA + \alpha \cdot (\cent(s)A)A$.  This shows that $\cent(s)A$ is by definition a centrality vector for starting vector $sA$.  Moreover, we know that the solution to $\cent(sA)$ is unique, so we have $\cent(s)A = \cent(sA)$.  This observation shows that we can iteratively compute the centrality vector by expressing $\cent(s)A$ as $\cent(sA)$.

We will write the operations performed inside the while-loop using vector-matrix notation.  We use $e_{u}$ to denote a row vector that has all of its content in vertex $u$:  $e_{u}(i)= 1$ if $i = u$; otherwise, $e_{u}(i)= 0$.

After an iteration of the loop we have  $\tilde{cr}' = \tilde{cr} + r(u)e_{u}$, and $r' = r - r(u)e_{u} + \alpha r(u)e_{u}A$, where $u$ is the vertex that is dequeued in line 11.  We next specify the relationship between the approximate centrality and residual vectors before and after an iteration of the while-loop.  Consider that

\begin{displaymath}
\begin{array}{lcl}
\cent(r)& = & \cent(r - r(u)e_{u}) + \cent(r(u)e_{u})\\
& = & \cent(r - r(u)e_{u}) + r(u)e_{u} + \cent(\alpha r(u)e_{u}A)\\
& = & \cent(r - r(u)e_{u} + \alpha r(u)e_{u}A) + r(u)e_{u}\\
& = & \cent(r') + \tilde{cr}' - \tilde{cr}.
\end{array}
\end{displaymath}

If $\tilde{cr} = \cent(s - r)$, we have $\cent(r) = \cent(r') + \tilde{cr}' - \cent(s-r)$.  It follows that $\tilde{cr}' = \cent(r) - \cent(r') + \cent(s-r) = \cent(r - r' + (s-r)) = \cent(s-r')$.
\end{proof}

\begin{theorem}\label{theorem:Correctness}
Given an $\alpha \le \frac{c}{d_{\max}}$ for some $c < 1$ and a uniform starting vector $s$, the vector $\tilde{cr}$ output by \emph{Approximate-Centrality} satisfies $\lbrack \cent(s) \rbrack (u) \ge \tilde{cr}(u) \ge \lbrack \cent(s) \rbrack (u)(1-\delta)$ for each vertex $u \in V$.  The runtime of the algorithm is $O(\frac{n}{\delta}d_{max})$.
\end{theorem}

\begin{proof}
Lemma~\ref{lemma:LoopInvariant} argues that $\tilde{cr} = \cent(s - r) = \cent(s) - \cent(r)$ throughout the execution of the algorithm, so we have $\tilde{cr}(u) = \lbrack \cent(s) \rbrack (u) - \lbrack \cent(r) \rbrack (u)$ for all vertices $u \in V$.  Given a uniform starting vector $s$, $s(u) = || s || _{1} / n$ for all
$u \in V$.  The algorithm terminates when $r(u) \le \epsilon$ for all $u \in V$, so we choose $\epsilon = \delta \cdot || s || _{1} / n = \delta s(u)$ such that upon completion $r(u) \le \delta s(u)$ for all $u \in V$.

Clearly, $\lbrack \cent(s) \rbrack (u) \ge \tilde{cr}(u)$ because $r$ and $\cent(r)$ are non-negative.  We can also show that given that $r(u) \le \delta s(u)$ for all $u \in V$, $\lbrack \cent(r) \rbrack (u) \le \delta \lbrack \cent(s) \rbrack (u)$ for all vertices $u \in V$.  It follows that $\tilde{cr}(u) = \lbrack \cent(s) \rbrack (u) - \lbrack \cent(r) \rbrack (u) \ge \lbrack \cent(s) \rbrack (u) (1-\delta)$.  Therefore we can see that indeed $\lbrack \cent(s) \rbrack (u) \ge \tilde{cr}(u) \ge \lbrack \cent(s) \rbrack (u)(1-\delta)$ for all vertices $u \in V$.

We assume that $\alpha$ is chosen such that $\alpha \le \frac{c}{d_{\max}}$ for some constant $c < 1$, where $d_{\max}$ is the largest out-degree of any node in the graph.  In order to bound the runtime of the algorithm, consider that each iteration of the while-loop decreases the sum of the entries of $r$ by $(1 - \alpha \cdot d_{\out}(u)) r(u) > (1 - \alpha \cdot d_{\out}(u)) \epsilon \ge (1 - \alpha \cdot d_{\max}) \epsilon \ge (1 - c) \epsilon$.  Because $r = s$ at initialization and each iteration decreases $|| r || _{1}$ by at least $(1-c)\epsilon$, the number of iterations $i$ must satisfy $i (1-c) \epsilon \le || s || _{1}$.  Therefore the number of iterations may be at most $\frac{|| s || _{1}}{(1-c)\epsilon} = O(|| s || _{1} / \epsilon)$.  The cost of each iteration is proportional to the out-degree of the node that is dequeued, so the worst-case runtime of the algorithm is $O(|| s || _{1} / \epsilon \cdot d_{\max})$.  For our choice of $\epsilon$ this is equivalent to $O(\frac{n}{\delta}d_{\max})$.
\end{proof}

\subsection{Quality of Approximate Results}
We compare the performance of the approximate algorithm with the power iteration method in Equation \ref{a-cen} using the indegree as the starting vector,  like in ~\cite{Bonacich87} and \cite{Bonacich01}. To compute Alpha-centrality using the approximate algorithm, we fix $\epsilon$ (Algorithm \ref{alg-main}) to be $3.57\times 10^{-8}$   and  $1.42\times 10^{-8}$ guaranteeing that the error in approximation would be less than $1\%$($\delta<0.01$).  We terminate the power iteration algorithm after 100 iterations in Digg and 10 to 100 iterations in Twitter. We calculate the RMS(root mean square) error of the approximate algorithm with respect to the power iteration algorithm, for different values of $\alpha$. The RMS error averaged over all values of $\alpha$, is 0.797\% and 0.75\% for Digg and Twitter respectively.
\section{Related Work}

The interplay of the structural properties of the underlying network with the diffusion processes occurring in it, contributes to the complexity of real-life networks.
For example in epidemiology, the dynamics of disease spread on a network and the epidemic threshold is closely related to its spectral radius of the graph ~\cite{Wang03}. Similarly, random walk on a graph is closely related Laplacian of the graph \cite{Chung_pagerankand}.

The  range of  diffusion processes that can occur on a network includes the spread of epidemics~\cite{Anderson91,Hethcote00} and information ~\cite{Lerman10icwsm}, viral  marketing~\cite{Kempe03,Iribarren09}, word-of-mouth recommendation~\cite{Goldenberg01}, money exchange, e-mail forwarding~\cite{Liben-Nowell08pnas}, and Web surfing~\cite{PageRank}, among others. Researchers have developed an arsenal of centrality metrics to study the properties of networks, including degree, closeness~\cite{sabidussi:1966},  graph~\cite{hage:1995} and betweenness~\cite{Freeman79};  Markov process-based random measures like the Hubbels model~\cite{Hubbell:1965}; path-based ranking measures like the Katz score~\cite{Katz53}, SenderRank~\cite{Kiss08}, and eigenvector centrality~\cite{Bonacich01}.
However, as Borgatti noted~\cite{Borgatti05}, most centrality measures make implicit assumptions about the diffusion process occurring on a network. In order to give correct predictions, these assumptions must match the actual dynamics of the network. Borgatti classified dynamic processes according to the trajectories they follow (geodesic, path, trail, walk) and the method of spread (transfer, serial or parallel duplication). We on the other hand maintain that a simpler classification scheme, that divides dynamic processes into conservative and non-conservative, captures the essential differences between them and informs the choice of the centrality metric.  Apart from PageRank and Alpha-Centrality, other measures can be classified as conservative or non-conservative.

Online social networks provide us the unique opportunity to study the dynamic processes occurring on networks.  Some studies compared empirical measures, such as tweets and mentions on Twitter~\cite{Cha10icwsm,Lee10}, with centrality metrics including PageRank and in-degree centrality. We on the other hand, differentiate between the two distinct methods of quantifying influence: \emph{estimating} influence by measuring dynamics of social network behavior and using centrality metrics to \emph{predict} influence. In addition, we evaluate the predictive influence models using the empirical measurements.

Similar to personalized PageRank~\cite{Jeh03} for conservative diffusion, each user's unique notion of importance in non-conservative diffusion can be captured using customized starting vector for individual users in Alpha-Centrality, leading to personalized Alpha-Centrality.  The use of residual vectors and incremental computation  in the calculation of approximate Alpha-Centrality leads to scalability of the method.  Moreover, as in personalized PageRank, these residual vectors can be  shared across multiple personalized views, scaling the personalized Alpha-Centrality metric.  Analogous to approximate PageRank~\cite{approximatePR}, in approximate Alpha-Centrality, at each iteration  residual vector is redistributed to reduce the difference between the Alpha-Centrality vector and its approximate version.  However, the process of redistribution of the residual vector mimics the kind of diffusion the model emulates. For approximate in PageRank, the redistribution of residual vectors is conservative (with the total weight of the residual vector conserved). On the other hand, in approximate Alpha-Centrality, the redistribution of residual vectors is not conservative.
\remove{
Since the formulation of  Alpha-Centrality is very similar to that of PageRank, similar block based strategies can be used to further speed up the computation of both PageRank and normalized Alpha-Centrality~\cite{Haveliwala:1999,Kamvar:2003}.
Since both Page Rank and alpha-centrality use Power Method to compute successive iterates, approaches to accelerate the convergence of the Power Method \cite{Brezinski,Kamvar03extrapolationmethods} can  be used to further speed-up Page Rank and alpha-centrality computations.
}

\section{Conclusion}
We described two fundamentally distinct diffusion processes, which can be mathematically differentiated based on whether or not they conserve the quantity that is diffusing on the network. Random walk, which conserves the probability density of the diffusing quantity, can be modeled as a conservative diffusion process, while epidemics and information spread can be modeled as non-conservative diffusion process. 
We showed that centrality metrics, such as PageRank and Alpha-Centrality, can be classified as conservative or non-conservative based on the implicit assumptions they make about the redistribution of weight. We showed that since Alpha-Centrality is mathematically equivalent to non-conservative diffusion, it should be used to identify central nodes in online social networks whose primary function is to spread information, a non-conservative process.
Future work includes applying this analysis to other online social networks like Twitter and  exploring how diffusion process affect other aspects of social network analysis. Our work provides just the initial study of non-conservative diffusion --- much work has to be done to understand its properties and extension, for example, application to personalized Alpha-Centrality may be productive.
We hope that our work motivates readers to study the properties of non-conservative diffusion and investigate the use of non-conservative in social network analysis.

\bibliographystyle{IEEEtran}
\bibliography{references}

\section*{Appendix}
\label{ch:Appendix}
Replication matrix $\WN$ can be written  in terms of its eigenvalues and eigenvectors as:
\begin{equation}
\WN=X\Lambda X^{-1} = \sum_{i=1}^{|V|} \lambda_{i}Y_{i}
\label{eq:eigen}
\end{equation}
where $X$ is a matrix whose columns are the eigenvectors of $\WN$.
$\Lambda$ is a diagonal matrix, whose diagonal elements are the eigenvalues, $\Lambda_{ii}=\lambda_{i}$, arranged according to the ordering of the eigenvectors in $X$.
Without loss of generality we assume that $\lambda_{1}> \lambda_{2} >\cdots >\lambda_{n}$.
The matrices $Y_{i}$ can be determined from the product
\begin{equation}
Y_{i}=X {Z}_{i}X^{-1}
\label{eq:z}
\end{equation}
where $Z_{i}$ is the \emph{selection matrix} having zeros everywhere except for element ${(Z_i)}_{ii}=1$ ~\cite{Gebali:2008}. 
Therefore 
\begin{eqnarray}
S({\alpha,t})&= &{\sum_{k=0}^{t}  (\alpha \WN)^k} \nonumber \\
&=& I+ \alpha \lambda_1{\displaystyle \sum_{i=1}^n} \frac{{(-1)}^{\mathcal{I}_i}(1-\alpha^{t+1} \lambda_{i}^{t+1})}{{(-1)}^{\mathcal{I}_i}(1-\alpha \lambda_{i})} Y_{i}
\label{eq:cm}
\end{eqnarray}
where $\mathcal{I}_i=0$ if $\alpha \left |\lambda_{i}\right | <1$ and $\mathcal{I}_i=1$ if $\alpha \left |\lambda_{i}\right | >1$. As obvious from above, for  Equation \ref{eq:cm}  to hold non-trivially, $\alpha \neq \frac{1}{\left |\lambda_{i}\right|} \forall i \in 1,2\cdots,n$. Now assuming $|\lambda_{1}|$
    is strictly greater than any other eigenvalue 
    $$S({\alpha,t}) \approx  I+ \frac{{(-1)}^{\mathcal{I}_1}(\alpha \lambda_1(1-\alpha^{t+1} \lambda_{1}^{t+1}))}{{(-1)}^{\mathcal{I}_1}(1-\alpha \lambda_{1})} Y_{1}.$$

  For any matrix $M$, let $||M||_1= \sum_{i,j} M[i,j]$  Therefore, the expected number of paths is $||S(\alpha,t)||_1$.    
  The expected path length is given by:
\begin{eqnarray*}
 \frac{{\displaystyle \sum_{k=0}^t} k\alpha^{k}||\WN^{k}||_1}{{\displaystyle \sum_{k=0}^t} \alpha^{k}||\WN^{k}||_1} &=& \frac{\alpha \frac{d||S(\alpha,t)||_1}{d\alpha}}{||S(\alpha,t)||_1} \nonumber  \\
&\approx & {{(-1)}^{\mathcal{I}_i}  (\frac{1}{1-\alpha \lambda_1} - (t+1) \frac{\alpha^{t+1}\lambda_1^{t+1}}{1- \alpha^{t+1}\lambda^{t+1}})} 
\label{eq:cm1}
\end{eqnarray*}
Therefore,  as $t\to \infty$ and $\alpha|\lambda_1|<1$, the expected path length is approximately  $\frac{1}{1-\alpha \lambda_1}$, and for $\alpha|\lambda_1|>1$ it is $O(t)$.

\end{document}